\documentclass[twocolumn,showpacs,preprintnumbers,amsmath,amssymb]{revtex4}

\usepackage{graphicx}
\usepackage{dcolumn}
\usepackage{bm}
\usepackage{amsmath}
\usepackage{slashed}
\usepackage{latexsym}

\newcommand{\bq}{\begin{equation}}
\newcommand{\eq}{\end{equation}}
\newcommand{\bqa}{\begin{eqnarray}}
\newcommand{\eqa}{\end{eqnarray}}
\newcommand{\nn}{\nonumber \\}

\def\be     {\begin{equation}}
\def\ee     {\end{equation}}
\def\bea        {\begin{eqnarray}}
\def\eea        {\end{eqnarray}}
\def\bnn    {\begin{eqnarray*}}
\def\enn    {\end{eqnarray*}}

\begin{document}

\title{Interplay between interaction and chiral anomaly: Anisotropy in the electrical resistivity of interacting Weyl metals}
\author{Yong-Soo Jho$^{1}$ and Ki-Seok Kim$^{1,2}$}
\affiliation{ $^{1}$Department of Physics, POSTECH, Hyoja-dong,
Namgu, Pohang, Gyeongbuk 790-784, Korea \\ $^{2}$Institute of Edge
of Theoretical Science (IES), Hogil Kim Memorial building 5th
floor, POSTECH, Hyoja-dong, Namgu, Pohang, Gyeongbuk 790-784,
Korea }
\date{\today}

\begin{abstract}
We predict that long-range interactions give rise to anisotropy in the electrical resistivity of Weyl metals at low temperatures, where 
the electrical resistivity becomes much reduced when electric fields are applied to the direction of the momentum vector to connect two paired Weyl points. 
Performing the renormalization group analysis, we find that the distance between two Weyl points becomes enhanced logarithmically at low temperatures 
although the coupling constant of such interactions vanishes inverse-logarithmically. Considering the Adler-Bell-Jackiw anomaly, scattering between these two Weyl
points becomes suppressed to increase electrical conductivity in the ``longitudinal" direction, counter-intuitive in the respect that interactions are expected 
to reduce metallicity. We also propose that the anomalous contribution in the Hall effect shows the logarithmic enhancement as a function of temperature,
originating from the fact that the anomalous Hall coefficient turns out to be proportional to the distance between two paired Weyl points. Correlations
with topological constraints allow unexpected and exotic transport properties.
\end{abstract}


\maketitle

\section{Introduction}

Nontrivial global structures of ground states sometimes violate
classically respected conservation laws at quantum levels,
referred to as (quantum) anomalies \cite{Peskin}. When anomalies
are associated with breaking of local (gauge) symmetries, it means
that their corresponding quantum theories are not consistent and
such anomalies should be cancelled, introducing meaningful quantum
fields. Indeed, the standard model and string theories are
constructed consistently, cancelling gauge and gravitational (also
conformal) anomalies, respectively \cite{GSW_String}. On the other
hand, when such anomalies are related with breaking of global
symmetries, they give rise to various interesting physical
properties. In particular, various types of topological terms
associated with quantum anomalies arise to play essential roles in
quantum criticality of quantum matter \cite{Fendley_NLsM}. In
addition, they turn out to be responsible for quantum number
fractionalization, given by Goldstone-Wilczek currents
\cite{Review_FZM}. Actually, an emergent non-abelian chiral
anomaly has been proposed to cause so called deconfined quantum
criticality in low dimensional spin systems \cite{Senthil,Tanaka}.
Furthermore, such topological terms sometimes give rise to
anomalous (quantized) electrical or thermal Hall effects
\cite{Review_AHE,Review_TI}. Quantum anomalies govern quantum
criticality, quantum number fractionalization, and anomalous
transport phenomena \cite{Kim_AdS5}.

In this study we focus on the role of the Adler-Bell-Jackiw
anomaly or chiral anomaly \cite{Peskin} in anomalous transport
phenomena. This anomaly means that classically conserved chiral
currents, that is, currents of right-handed (Weyl) fermions minus
those of left-handed fermions, are not preserved at quantum levels
due to nontrivial global configurations of gauge (or electric and
magnetic) fields. Applying magnetic fields to gapless
semi-conductors, a Dirac point described by the four-component
Dirac spinor splits into two Weyl points governed by the
two-component Weyl spinors with opposite chiralities, where the
distance between two Weyl points is proportional to the applied
magnetic field \cite{Weyl_Metal}. The chiral anomaly gives rise to
a topological constraint in dynamics of Weyl fermions, where
right-handed Weyl fermions at one Weyl point should scatter into
left-handed Weyl fermions at the other Weyl point, when currents
are driven to the same direction as the momentum to connect these
two Weyl points \cite{Anomaly_Weyl}. Even if short-range
scatterers are taken into account, scattering between these two
Weyl points becomes suppressed due to the finite distance in the
momentum space. As a result, the longitudinal ($E
\parallel B$) magneto-conductivity is enhanced, which turns out to be
proportional to the square of the applied magnetic field or the
distance of two Weyl points \cite{Anomaly_Weyl,DTSon_Boltzmann}.

In this paper we investigate effects of interactions on the
``longitudinal" ``magneto"-transport in Weyl metals. Here, `` "
will be clarified later. It is almost trivial to observe that
local four-fermion interactions are irrelevant at low energies in
a perturbative sense since the density of states vanishes at zero
energy. Long-range Coulomb interactions have been investigated
both extensively and intensively for transport phenomena in
graphene \cite{Graphene_Review}. In addition, transverse gauge
interactions have been also discussed in Weyl- or Dirac-type
systems \cite{Nagaosa_Lee_Wen}. Recently, the chiral anomaly has
been calculated in the Weyl system \cite{Fujikawa_Method}.
However, the interplay between long-range interactions and the
chiral anomaly has not been investigated clearly. In particular,
it remains mysterious how this combination gives rise to anomalous
``longitudinal" ``magneto"-transport phenomena.

Performing the renormalization group analysis, we reveal that the
distance between two Weyl points becomes enhanced logarithmically
at low temperatures although the coupling constant for transverse
long-range interactions vanishes inverse-logarithmically (expected
in three dimensions). This is in contrast with ``conventional"
Weyl metals without interactions
\cite{Weyl_Metal,Anomaly_Weyl,DTSon_Boltzmann}, where the distance
between two corresponding Weyl points remains finite. As a result,
scattering between two Weyl points becomes suppressed much more
than the case of noninteracting Weyl metals, which increases
electrical conductivity in the direction to connect the momentum
vector between two Weyl points. We predict that anisotropic metallicity arises,
where the electrical resistivity becomes much reduced for the longitudinal direction
while normal metallic behaviors result for other directions. 
Furthermore, we propose that the anomalous contribution in the Hall effect 
becomes enhanced as a function of temperature, originating from the fact 
that the anomalous Hall coefficient turns out to be proportional to 
the distance between two paired Weyl points \cite{Ran_AHE_Weyl}. We discuss this interaction-enhanced 
anisotropy in the longitudinal resistivity and the increase of the anomalous contribution in the
Hall effect, based on the quantum Boltzmann equation approach 
in the presence of both long-range transverse interactions and the chiral anomaly.

\section{Interplay between long-range transverse interactions and the chiral anomaly}

We start from quantum electrodynamics with a topological
$\theta$-term in three spatial dimensions ($\theta-$QED$_{4}$)
\bqa && \mathcal{L} = -\frac{1}{4}F_{\mu \nu}F^{\mu\nu}+
i\bar{\psi} \slashed{D}  \psi +\frac{e^2\theta}{8 \pi^2} F_{\mu
\nu} \tilde{F}^{\mu \nu},  \eqa where $\psi$ is a four-component
Dirac spinor to take both chirality (associated with either
orbital or sublattice indices) and spin quantum numbers and
$A_{\mu}$ is an electromagnetic vector potential regarded as a
quantum field. $ D_\mu = \partial_\mu + i e A_\mu $ in $
\slashed{D} = \gamma^{\mu} D_{\mu} $ is a covariant derivative
with an electric charge $e$, where $\gamma^{\mu}$ is the Dirac
gamma matrix satisfying the Clifford algebra with $\mu = 0, 1, 2,
3$. $ \tilde{F}^{\mu \nu} = \frac{1}{2} \epsilon^{\mu \nu \rho
\sigma} F_{\rho \sigma} $ is a magnetic dual tensor of the
electromagnetic field strength tensor. One may consider that this
field theory results from certain lattice models with spin-orbit
interactions for topological insulators except for long-range
transverse interactions \cite{Chiral_Gauge_Field}. However, we
would like to emphasize that even gauge fluctuations can emerge in
some lattice models, supporting so called topological spin liquids
\cite{TSL_YB}. The $\theta-$term is the fingerprint of the
topological insulator in three dimensions, the source of the
(longitudinal) magnetoelectric effect or equivalently, the
half-quantized Hall conductance on its surface \cite{Review_TI}.

The Adler-Bell-Jackiw anomaly states that the classically
conserved chiral current is not conserved at quantum levels
\cite{Peskin}, given by \bqa &&  \partial_\mu J^{5 \mu } = -
\frac{e^2}{8 \pi^2} F_{\mu \nu} \tilde{F}^{\mu \nu} \eqa with the
chiral current $J_{\mu}^{5} = \bar{\psi} \gamma_{\mu} \gamma_{5}
\psi = J_{\mu}^{R} - J_{\mu}^{L}$ mentioned before, where the
$\gamma_{5}$ matrix is a Dirac matrix that anti-commutes with
other Dirac matrices. In other words, when electric fields are
applied in parallel with magnetic fields, the chiral current is
not conserved. It is important to realize that the $\theta-$term
is a boundary term, implying that the coefficient $\theta$ cannot
be renormalized by interactions. However, if inhomogeneous
magnetic fields can be applied to this topological insulating
state, the $\theta$ coefficient depends on position
\cite{Chiral_Gauge_Field,Axion_Insulator}. As a result, this term
is not a boundary term any more, which can be renormalized by
interactions.

Resorting to this anomaly equation, we rewrite the effective field
theory as follows \bqa && \mathcal{L} =
-\frac{1}{4}F_{\mu \nu}F^{\mu\nu}+  \bar{\psi} (
i\slashed{\partial} - e \slashed{A} + \slashed{c} \gamma_5 )  \psi
,  \eqa where $c_{\mu}$ with the $\gamma_{5}$ Dirac matrix is a
chiral gauge field, given by $\partial_{\mu} \theta$. See appendix A1
for the derivation from Eq. (3) to Eq. (1) with Eq. (2).
It is interesting to observe that the Dirac point splits into two Weyl
points at $\boldsymbol{K} = \pm \boldsymbol{c}$. In this respect
our problem is to investigate the nature of the quantum critical
point between a topological insulator and a band insulator in the
presence of inhomogeneous magnetic fields. In other words, we
study how both the interaction parameter $e$ and the distance
between two Weyl points $c_{\mu}$ are renormalized to evolve at
low temperatures, and reveal how these renormalization effects
modify transport properties, compared with the case in the absence
of interactions. Such inhomogeneous magnetic fields may be created
by either ferri-magnetism \cite{Chiral_Gauge_Field} or some
ferromagnetic clusters, given by randomly distributed magnetic
ions \cite{MI_TI_Kim}.

Introducing counter terms, we rewrite this effective field theory
as follows \bqa && \mathcal{L} =
-\frac{Z_A}{4}F_{\mu \nu}F^{\mu\nu} +  \bar{\psi} \left ( Z_\psi i
\slashed{\partial} + Z_c \slashed{c} \gamma_5 - Z_e e \slashed{A}
\right ) \psi   , \nonumber \\ && \eqa where $Z_{\psi}$, $Z_{A}$,
$Z_{c}$, and $Z_{e}$ are field renormalization constants of
$\psi$, $A_{\mu}$, $c_{\mu}$ and the vertex or coupling ($e$)
renormalization constant, respectively. See appendix A1 for details. We emphasize that the
renormalization factor $Z_{c}$ has never been introduced as far as
we know, thus regarded as an essential aspect of this study. We
alert that $c_{\mu}$ is a background gauge field, not dynamical in
the present study.

In order to perform the renormalization group analysis, we resort
to the dimensional regularization. A subtle point arises due to
the presence of the $\gamma_{5}$ matrix, which needs some care for
its treatment, because its existence depends on dimensionality
\cite{Peskin}. We point out that the presence of the $\gamma_{5}$
matrix makes our calculations much more complicated and laborious.
One nontrivial check for the validity of our calculations is that
the Ward identity is respected in the one-loop level. All details
are presented in appendix A2.
As a result, we obtain our coupled renormalization group equations
for $e$ and $c_{\mu}$, \bqa && \beta_e (\mu) = \mu \frac{d e}{d
\mu}= \frac{e^3}{12 \pi^2}, \\ && \beta_c (\mu) = \mu\frac{d
c_\nu}{d \mu} = - \frac{e^2}{4 \pi^2}c_\nu . \eqa See appendix A3
doe the derivation of Eqs. (5) and (6).

The first equation is nothing but the conventional renormalization
group equation of the coupling constant, which tells us that the
electric charge renormalizes to vanish at zero temperature. Even
if the coupling constant vanishes, the background chiral gauge
field flows to go to infinity. Inserting the solution of the first
equation into the second equation, we find 
\bqa &&  e^2(T) = \frac{e_D^2}{ 1 + \frac{e_D^2}{4 \pi^2} \ln
\left ( \frac{D}{T} \right ) }  , \\ && c_\nu (T) = c_\nu^{D} \
\left | 1 + \frac{e_D^2}{4 \pi^2} \ln \frac{D}{T} \right |, \eqa
where $c_{\mu}^{D}$ and $e_{D}$ are chiral gauge field and
coupling constant at the energy scale of the bandwidth or cutoff. See appendix A4.
The chiral gauge field increases in a logarithmic way. This
indicates that the distance between two Weyl points becomes
``infinite" at zero temperature, implying that their scattering
events are suppressed ``completely". This renormalization effect
should be observed in transport coefficients.

\section{Anisotropy in the longitudinal electrical transport
and enhancement of the anomalous contribution in the Hall effect}

Our framework for anomalous transport is the ``semi-classical"
quantum Boltzmann equation approach. Here, the term
``semi-classical" means that the role of both Berry curvature and
chiral anomaly or the topological $\theta-$term is introduced from
coupled semi-classical equations of motion based on the
wave-packet picture in solids \cite{Review_AHE}. Benchmarking a recent 
transport study based on the classical Boltzmann equation \cite{DTSon_Boltzmann},
we incorporate this information into the quantum Boltzmann equation, 
which have been applied to transport dynamics in strongly correlated electrons 
\cite{Kim_QBE}. As a result, inelastic scattering events can be taken into account naturally in
the presence of the topological $\theta-$term. 
We consider the case of a finite chemical potential, more generic
than the case with two Weyl points. In principle, one can derive
the quantum Boltzmann equation in a matrix form, regarded as a
full quantum transport theory \cite{Matrix_Boltzmann_Equation,Shindou_QBE}.
However, its derivation is much complicated and not easy to
perform. We would like to emphasize that our phenomenological
``quantum" transport theory with the introduction of the topological $\theta-$term
recovers the known result for the longitudinal transport coefficient in Weyl metals, 
the so called ``negative magnetoresistance" proposed in Ref. \cite{Anomaly_Weyl}.  

We start from the quantum Boltzmann equation for a steady state
\cite{Mahan_Boltzmann} \bqa && \boldsymbol{\dot{p}} \cdot
\frac{\partial G^{<}(\boldsymbol{p},\omega)}{\partial
\boldsymbol{p}} + \boldsymbol{\dot{r}} \cdot \boldsymbol{\dot{p}}
\frac{\partial G^{<}(\boldsymbol{p},\omega)}{\partial \omega} \nn && -
\boldsymbol{\dot{p}} \cdot \Bigl\{ \frac{\partial
\Sigma^{<}(\boldsymbol{p},\omega)}{\partial \omega} \frac{\partial
\Re G_{ret}(\boldsymbol{p},\omega)}{\partial \boldsymbol{p}} -
\frac{\partial \Re G_{ret}(\boldsymbol{p},\omega)}{\partial
\omega} \frac{\partial \Sigma^{<}(\boldsymbol{p},\omega)}{\partial
\boldsymbol{p}} \Bigr\} \nn && = - 2 \Gamma(\boldsymbol{p},\omega)
G^{<}(\boldsymbol{p},\omega) + \Sigma^{<}(\boldsymbol{p},\omega)
A(\boldsymbol{p},\omega) . \eqa $G^{<}(\boldsymbol{p},\omega)$ is
the lesser Green's function, regarded as a quantum distribution
function, where $\boldsymbol{p}$ and $\omega$ represent momentum
and frequency for relative coordinates, respectively.
$\dot{\mathcal{O}}$ denotes the derivative with respect to time
$t$. $\Sigma^{<}(\boldsymbol{p},\omega)$ and
$G_{ret}(\boldsymbol{p},\omega)$ indicate the lesser self-energy
and the retarded Green's function, respectively, where $\Re$ is
their real part. The right hand side introduces collision terms,
where $\Gamma(\boldsymbol{p},\omega)$ and
$A(\boldsymbol{p},\omega)$ indicate the scattering rate and the
spectral function.

$\boldsymbol{r}$ and $\boldsymbol{p}$ are governed by
semi-classical equations of motion \cite{Review_AHE}, given by
\bqa && \boldsymbol{\dot{r}} = \frac{\partial
\epsilon_{\boldsymbol{p}}}{\partial \boldsymbol{p}} +
\boldsymbol{\dot{p}} \times \boldsymbol{\Omega}_{\boldsymbol{p}} ,
\nn && \boldsymbol{\dot{p}} = e \boldsymbol{E} + \frac{e}{c}
\boldsymbol{\dot{r}} \times \boldsymbol{B} , \eqa where
$\boldsymbol{\Omega}_{\boldsymbol{p}}$ represents the Berry
curvature of the momentum space. Solving these equations, one
obtains \bqa && \boldsymbol{\dot{r}} = \Bigl( 1 + \frac{e}{c}
\boldsymbol{B} \cdot \boldsymbol{\Omega}_{\boldsymbol{p}}
\Bigr)^{-1} \Bigl\{ \boldsymbol{v}_{\boldsymbol{p}} + e
\boldsymbol{E} \times \boldsymbol{\Omega}_{\boldsymbol{p}} +
\frac{e}{c} \boldsymbol{\Omega}_{\boldsymbol{p}} \cdot
\boldsymbol{v}_{\boldsymbol{p}} \boldsymbol{B} \Bigr\} , \nn &&
\boldsymbol{\dot{p}} = \Bigl( 1 + \frac{e}{c} \boldsymbol{B} \cdot
\boldsymbol{\Omega}_{\boldsymbol{p}} \Bigr)^{-1} \Bigl\{ e
\boldsymbol{E} + \frac{e}{c} \boldsymbol{v}_{\boldsymbol{p}}
\times \boldsymbol{B} + \frac{e^{2}}{c} (\boldsymbol{E} \cdot
\boldsymbol{B}) \boldsymbol{\Omega}_{\boldsymbol{p}} \Bigr\} . \nn
\eqa An essential point is the presence of the $\boldsymbol{E}
\cdot \boldsymbol{B}$ term in the second equation, imposing the
Adler-Bell-Jackiw anomaly \cite{DTSon_Boltzmann}.

Inserting Eq. (11) into Eq. (9) and performing straightforward but 
rather tedious algebra, we reach the following expression for the longitudinal
conductivity \bqa && \sigma_{L}(T)
\longrightarrow (1 + \mathcal{K} [c(T)]^{2}) \sigma_{n}(T) , \eqa
where $\mathcal{K}$ is a positive numerical constant and the
normal conductivity $\sigma_{n}(T)$ is determined by
(gauge-interaction induced) intra-scattering events at one Weyl
point. All details are shown in appendix B. 
An essential point is in the $c(T)$ term, where $c(T)$ is
the distance between two paired Weyl points, given by our renormalization
group analysis. Here, the vector index $\mu$ is fixed and omitted
for simplicity. Actually, this expression is to replace the
applied magnetic field in noninteracting Weyl metals with the
distance between Weyl points in interacting Weyl metals, when
electric fields are applied in parallel with ``magnetic fields" or
the momentum vector to connect such Weyl points, implying the
reason why we call ``longitudinal" in front of the conductivity.
We emphasize that this expression recovers that of the original
proposal \cite{Anomaly_Weyl}, where the magneto-conductivity is proportional to the
square of the distance between two Weyl points. Although the
distance does not renormalize in the noninteracting case, 
the presence of transverse long-range interactions gives rise
to the logarithmic enhancement at zero chemical potential. 

One aspect should be pointed out carefully. We proved that the
distance between two paired Weyl points increases logarithmically
at low temperatures, originating from long-range
transverse interactions. It should be noticed that this result
appears at zero chemical potential. On the other hand, we found
the longitudinal conductivity [Eq. (9)] for a finite chemical
potential. Can we expect the similar enhancement of the distance
between two paired Weyl points in the case of a finite chemical
potential? The renormalization group analysis for the Weyl metallic state
with a finite chemical potential turns out to be more complex and
technically involved, originating from the treatment of four by four matrices 
and angular integrals along Fermi surfaces. Besides such technical difficulties, 
we reach the conclusion that an exotic phenomenon may appear. It is natural to expect that
the Weyl metallic state with two corresponding Fermi surfaces will not be stable at low temperatures when there exist interactions.
This instability originates from their perfect nesting. As a result, some types of charge or spin density waves
are expected to arise at low temperatures. However, the main result of the previous section in the case of zero chemical potential
is that the distance between two paired Weyl points increases to diverge, implying that
scattering between these two Weyl points is suppressed. Then, we expect that
the competition between two kinds of divergences, one of which comes from the perfect nesting while the other of which
originates from the chiral anomaly with long-range transverse interactions, may allow
an exotic balance, which gives rise to an interacting fixed point, identified with a novel non-Fermi liquid metallic state.
In this respect we believe that the study in the case of a finite chemical potential should be
performed more carefully and would like to leave it as a future work. 

Although we cannot determine the temperature dependence of the distance between two Weyl points 
in the presence of Fermi surfaces, the presence of the prefactor $(1 + \mathcal{K} [c(T)]^{2})$ 
in the longitudinal resistivity guarantees an anisotropic metallic behavior because
such a prefactor does not exist in the transverse resistivity. Unfortunately, we cannot 
quantify the degree of anisotropy at present.

We propose another fingerprint of the interacting or critical Weyl metallic state with long-range transverse interactions,
that is, peculiar temperature dependencies for the anomalous contribution of the Hall coefficient, depending on the chemical potential. 
Recently, the anomalous contribution for the Hall effect has been evaluated \cite{Fujikawa_Method,Ran_AHE_Weyl}, given by 
\bqa \sigma_{\mu\nu} = \frac{e^{2}}{2\pi^{2}} \epsilon_{\mu\nu\gamma} \boldsymbol{c}_{\gamma} \eqa 
in the absence of Fermi surfaces, where $\boldsymbol{c}_{\gamma}$ is the distance between two paired Weyl points. 
Based on our renormalization group analysis [Eq. (8)], we find \bqa && \sigma_{xy}(T)  
= \frac{e^{2}}{2\pi^{2}} c_z(D) \ \left | 1 + \frac{e_D^2}{4 \pi^2} \ln \frac{D}{T} \right |  , \eqa where $c_z(D)$
is the distance of two paired Weyl points at $T = D$. This expression is rather unexpected because the anomalous Hall coefficient 
diverges, dominating over the normal Hall effect. Here, the term ``divergence" should be regarded more carefully. Since 
our long-wave-length effective description is valid only below the momentum cutoff, at most within the first Brillouin zone,
the term ``divergence" is more accurate to be replaced with enhancement at low temperatures.

We show $\sigma_{xy}(T)/\sigma_{xy}(D)$ with $\sigma_{xy}(D) = \frac{e^{2}}{2\pi^{2}} c_z(D)$ in Fig. 1 for clarity of physics.
Interestingly, it has been also shown that Eq. (13) is not modified even in the presence of a finite chemical potential, based on the Kubo formula 
\cite{Fujikawa_Method}. It will be quite interesting to reveal the temperature dependence for the anomalous Hall coefficient
in the presence of Fermi surfaces near two paired Weyl points.

\begin{figure}[t]
\includegraphics[width=0.45\textwidth]{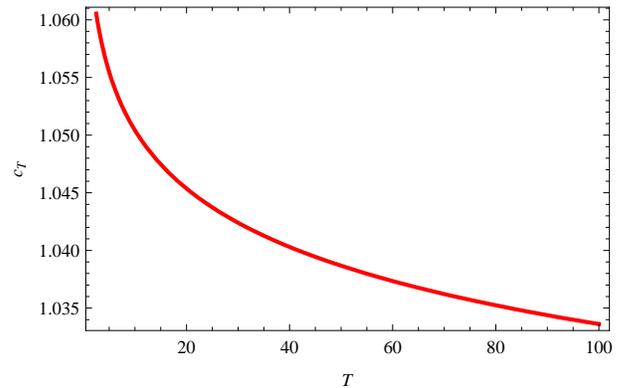}
\caption{Temperature dependence of the anomalous contribution for the Hall effect 
in an interacting Weyl metal with zero chemical potential. 
We plot a dimensionless anomalous Hall coefficient $\sigma_{xy}(T)/\sigma_{xy}(D)
= \left | 1 + \frac{e_D^2}{4 \pi^2} \ln \frac{D}{T} \right |$ as a function
of temperature $T$ of a unit of $K$, using $D = 10^{4} K$ and 
$\frac{e_{D}^{2}}{4\pi^{2}} = 1/137$. An essential feature is the 
logarithmic enhancement of the anomalous Hall coefficient as
a result of the interplay between transverse long-range
interactions and the chiral anomaly.} \label{fig1}
\end{figure}

\section{Conclusion and perspectives}

In summary, we investigated the quantum critical point of the
topological phase transition from a topological insulator to a
band insulator in the presence of inhomogeneous ferromagnetism or
under nonuniform magnetic fields, where the topological
$\theta-$term gives rise to a topological constraint in dynamics
of bulk fermions, referred to as the Adler-Bell-Jackiw anomaly.
Such inhomogeneous magnetic fields serve background chiral gauge
fields, splitting the Dirac point into two Weyl points.
Introducing long-range transverse interactions and performing the
renormalization group analysis, we uncovered that the distance
between these two Weyl points becomes enhanced logarithmically
at low temperatures although the coupling constant vanishes as
expected. Resorting to the semi-classical quantum Boltzmann
equation approach, we claimed that the enhancement of the distance
strengthens metallic properties at low temperatures when electric
fields are applied to the same direction as the momentum to
connect these Weyl points because scattering between the Weyl
points are suppressed due to their huge distance in the momentum
space. Besides this emergent enhanced anisotropy in electrical resistivity,
we predicted the logarithmically ``divergent" temperature dependence 
for the anomalous contribution of the Hall effect. These two anomalous 
transport properties are proposed to be fingerprints of Weyl metals
with transverse long-range interactions. 

There remain three important problems in our direction. The first question is to perform
the renormalization group analysis in the case of a finite chemical potential,
as discussed before. Since the competition between the enhancement of the
distance between two paired Weyl points and the presence of perfect nesting 
between two paired Fermi surfaces is expected to cause a delicate balance,
we are expecting an interacting fixed point, which can be identified with
a novel non-Fermi liquid metal. The second question is what
happens if we take into account chiral gauge fields quantum
mechanically. This situation arises when ferromagnetic phase
transitions occur near the topological phase transition
\cite{Chiral_Gauge_Field}. Is it possible to obtain a novel
interacting fixed point, too? The third question is more practical
thus experimentally verified. If we introduce weak
anti-localization corrections in the transport theory, how is the
longitudinal transport coefficient modified? This question is
still meaningful even without interactions because this transport
signature can be measured actually \cite{Weyl_LMR}.

KS was supported by the National Research Foundation of Korea
(NRF) grant funded by the Korea government (MEST) (No.
2012000550).

\begin{widetext}

\appendix

\section{Renormalization group analysis in the presence of chiral anomaly}

\subsection{Introduction of counter-terms}


We start from the following Lagrangian \bqa &&  \mathcal{L}_B = i
\bar{\psi}_B \gamma^\mu
\partial_\mu \psi_B - e_B A_{B\mu} \bar{\psi}_B \gamma^\mu \psi_B
- \frac{1}{4}F_{B \mu \nu} F_B^{\mu \nu} + \frac{e_B^2
\theta_B}{16\pi^2} \epsilon^{\mu \nu \rho \sigma} F_{B \mu \nu}
F_{B \rho \sigma} , \eqa where $ \psi_B, \ A_{B\mu}, \ e_B, \
\theta_B $ are bare quantities to be renormalized. Introducing
renormalization factors of $ \psi_B = Z_\psi^{1/2} \psi , \ A_{B
\mu} = Z_A^{1/2} A_{\mu}, \ e_B Z_A^{1/2} Z_\psi = Z_e e, \
\theta_B = Z_c \theta $, one can rewrite the above bare Lagrangian
in terms of its renormalized part and counter-term part, \bqa
&& \mathcal{L}_B = \mathcal{L}_r + \mathcal{L}_{c.t.} \nonumber \\
&& \mathcal{L}_r = i \bar{\psi} \gamma^\mu \partial_\mu \psi - e
A_\mu \bar{\psi} \gamma^\mu \psi - \frac{1}{4}F_{\mu \nu} F^{\mu
\nu} + \frac{e^2 \theta}{16 \pi^2}\epsilon^{\mu \nu \rho \sigma}
F_{\mu \nu} F_{\rho \sigma} \nonumber \\ && \mathcal{L}_{c.t.} =
\delta_\psi i \bar{\psi} \gamma^\mu \partial_\mu \psi - \delta_e e
A_\mu \bar{\psi} \gamma^\mu \psi - \frac{\delta_A}{4}F_{\mu \nu}
F^{\mu \nu} + \delta_c \frac{e^2 \theta}{16 \pi^2}\epsilon^{\mu
\nu \rho \sigma} F_{\mu \nu} F_{\rho \sigma},  \eqa where $
\delta_\psi = Z_\psi - 1 , \ \delta_e = Z_e - 1 , \ \delta_A = Z_A
- 1 , \ \delta_c = Z_c -1 $.

Incorporating the anomaly equation [Eq. (2)] for renormalized
fields into the above expression, we obtain Eq. (4) for our
renormalization group analysis. This procedure can be performed in
a more formal way. This Lagrangian functional is invariant under
the chiral transformation of $ \psi \to e^{i \alpha(x) \gamma_5 }
\psi $ as long as fermions remain massless. Consider the following
replacement \bqa && Z = \int \mathit{DA} \mathit{D\psi}
\mathit{D\bar{\psi}}\ \mathbf{exp} \left \{ i \int d^4 x
\mathcal{L}_r + \mathcal{L}_{c.t.} \right \} \nonumber \\ && {}
\to Z = \int \mathit{DA} \mathit{D\psi} \mathit{D\bar{\psi}}\
\mathbf{exp} \left \{ i \int d^4 x \mathcal{L}_r +
\mathcal{L}_{c.t.} + \alpha (x) \left (
\partial_\mu J^{\gamma_5 \mu} + \frac{e^2}{16 \pi^2} \epsilon^{\mu
\nu \rho \sigma} F_{\mu \nu } F_{\rho \sigma}  \right )  \right \}
, \eqa where gauge fixing is assumed. $ \alpha(x) $ is an
arbitrary and infinitesimal local parameter. Taking $ \alpha(x) =
- \theta(x) $, we see that the $ \theta F \tilde{F}$ term is
replaced with the chiral gauge-field term in Eq. (4). As a result,
we obtain the following expression \bqa && Z = \int \mathit{DA}
\mathit{D\psi} \mathit{D\bar{\psi}}\ \mathbf{exp} \left \{ i \int
d^4 x \mathcal{L}_1 + \mathcal{L}_{1.c.t.} \right \} \nonumber \\
&& \mathcal{L}_1 =  -\frac{1}{4}F_{\mu \nu}F^{\mu\nu}+
i\bar{\psi}\gamma^\mu (\partial_\mu -i c_\mu \gamma_5 )  \psi - e
A_\mu \bar{\psi} \gamma^\mu \psi \nonumber \\ &&
\mathcal{L}_{1.c.t.} =  -\frac{\delta_A}{4}F_{\mu \nu}F^{\mu\nu}+
i\bar{\psi}\gamma^\mu (\delta_\psi \partial_\mu -i \delta_c c_\mu
\gamma_5 )  \psi - \delta_e e A_\mu \bar{\psi} \gamma^\mu \psi ,
\eqa where gauge fixing is also assumed. It is important to notice
that the chiral anomaly equation is satisfied for renormalized
fields, not bare fields.

\subsection{One-loop structure of QED$_{4}$ with a background chiral gauge field}

We perform one-loop renormalization group analysis in the presence
of the background chiral gauge field, i.e., $ c_\mu $ = constant.
We obtain renormalization constants from $ \Sigma_1 (\slashed{p},
c), \ \Gamma_1^\mu (p, p', c) , \ \Pi_1^{\mu \nu} (q, c) $,
corresponding to one-loop fermion self-energy, one-loop vertex
correction, and one-loop gauge-boson self-energy, respectively.

It is important to notice that the background chiral gauge field
is taken into account non-perturbatively. In other words, our
vacuum state is a state that chiral currents are flowing. Thus,
the fermion propagator is modified to be \bqa &&
\frac{1}{\slashed{p}} \to \frac{1}{\slashed{p} +
\slashed{c}\gamma_5}= \frac{(p^2 + c^2 ) \slashed{p} + 2(p \cdot c
) \slashed{p} \gamma_5 - (p^2 + c^2 ) \slashed{c} \gamma_5 - 2 (p
\cdot c ) \slashed{c} }{(p-c)^2 (p+c)^2} . \eqa In this respect
the key point is how self-energies of fermions and gauge bosons
and vertex corrections are modified in this novel vacuum state. A
subtle point arises due to the presence of the $\gamma_{5}$ matrix
in the regularization procedure \cite{Peskin}. When dimensional
regularization is used to regularize loop-integrals including $
\gamma_5 $, some anomalous terms appear. They originate from
components perpendicular to physical four-dimensions. We separate
out these perpendicular momentum components of $ l_\perp^\mu =
l^\mu - l_\parallel^\mu $ explicitly in our dimensional
regularization. However, it turns out that they do not result in
divergent contributions.

First, we calculate the fermion self-energy \bqa && -i
\Sigma_1(\slashed{p},c) = (-ie)^2 \int \frac{d^d
k}{(2\pi)^d}\gamma^\mu i  \frac{ (k^2 + c^2 )\slashed{k} - 2 ( k
\cdot c ) \slashed{c} + 2 ( k \cdot c ) \slashed{k} \gamma_5 -
(k^2 + c^2)\slashed{c}\gamma_5  }{(k+c)^2(k-c)^2}\gamma^\nu
\frac{-i \eta_{\mu \nu}}{(p-k)^2} \nonumber \\ && {}= - e^2 \int
\frac{d^d k}{(2\pi)^d} \frac{1}{(p-k)^2 (k-c)^2 (k+c)^2 } \left \{
(k^2 + c^2) \gamma^\mu \slashed{k} \gamma_\mu - 2(k \cdot c)
\gamma^\mu \slashed{c} \gamma_\mu + 2 (k \cdot c)  \gamma^\mu
\slashed{k} \gamma_5 \gamma_\mu - (k^2 + c^2 ) \gamma^\mu
\slashed{c} \gamma_5 \gamma_\mu   \right \}. \nonumber \\ && \eqa
Replacing the denominator with Feynman parameters, we rewrite the
above expression as follows \bqa && -i \Sigma_1(\slashed{p}, c) =
-2e^2 \int_{0}^{1} dx dy dz \ \delta(x+y+z-1) \int \frac{d^d l}{(2
\pi)^d}\frac{1}{\left [ l^2 - \Delta \right ]^3} \nonumber \\ && \
\ \ \ \ \ \ \ \ \ \ \ \ \ \ \ \ \ \ \ \ \ \ \ \ \times \left \{
-2( ( l + a)^2 + c^2 ) ( \slashed{l} + \slashed{a} ) +4 ( l \cdot
c + a \cdot c ) \slashed{c} + 4(l \cdot c + a \cdot c)(\slashed{l}
+ \slashed{a}) \gamma_5 - 2 ( (l+a)^2 + c^2) \slashed{c} \gamma_5
\right \}, \nonumber \\ && \eqa where $ l_\mu = k_\mu - a_\mu , \
a_\mu = x p_\mu + (y-z) c_\mu , \ \Delta = (x^2 - x) p^2 + (y^2-y
+ z^2-z - 2yz)c^2 + (2xy -2xz)(p \cdot c). $ Note that the degree
of divergence in each term depends on only the power of redefined
loop momenta $ l$. Since we need to calculate only divergent
terms, we consider \bqa && -i \Sigma_1(\slashed{p}, c) = -2e^2
\int_{0}^{1} dx dy dz \ \delta(x+y+z-1) \int \frac{d^d l}{(2
\pi)^d}\frac{1}{\left [ l^2 - \Delta \right ]^3} \nonumber \\ && \
\ \ \ \ \ \ \ \ \ \ \ \ \ \ \ \ \ \ \ \ \ \ \ \ \ \ \ \ \  \ \
\times \left \{ -2 l^2 \slashed{a} - 4 l^\lambda l^\kappa
a_\lambda \gamma_\kappa + 4 l^\lambda l^\kappa c_\lambda
\gamma_\kappa \gamma_5 - 2 l^2 \slashed{c} \gamma_5  \right \} +
finite. \nonumber \\ && \ \ \ \ \ \ \ \ \ \ \ \ \ = -2e^2
\int_{0}^{1} dx dy dz \ \delta(x+y+z-1) \times \nonumber \\ && {}\
\ \ \ \ \ \ \ \ \ \ \ \ \ \ \ \ \ \ \ \ \ \ \ \ \ \ \ \ \ \left \{
-2 ( \slashed{a} + \slashed{c} \gamma_5 ) \left ( \int \frac{d^d
l}{(2\pi)^d}\frac{l^2}{[l^2 - \Delta]^3} \right ) - 4(a_\lambda
\gamma_\kappa - c_\lambda \gamma_\kappa \gamma_5 ) \left ( \int
\frac{d^d l}{(2\pi)^d}\frac{l^\lambda l^\kappa}{[l^2 - \Delta]^3}
\right )   \right \} + finite. \nonumber \\ && \eqa Integrating
over the loop momenta $ l$, we obtain \bqa &&
-i\Sigma_1(\slashed{p}, c) = \frac{e^2 i}{\pi^2 \epsilon}
\int_{0}^{1} dx dy dz \ \delta (x+y+z-1) \left \{ \frac{3}{4}
\slashed{a} + \frac{1}{4} \slashed{c} \gamma_5 \right \} + finite
\nonumber \\ && {} \ \ \ \ \ \ \ \ \ \ \ \ = \frac{e^2 i}{8 \pi^2
\epsilon} ( \slashed{p} + \slashed{c} \gamma_5) + finite \eqa with
$ \epsilon = 4-d$.

Second, we calculate the vertex correction in the same way as the
above \bqa && \Gamma_1^\mu (p, p', c) = (-i e)^2 \int \frac{d^d
k}{(2\pi)^d} \left \{ \gamma^\nu i \frac{  (k'^2 + c^2 )
\slashed{k' } - 2(k' \cdot c) \slashed{c} + 2(k' \cdot c)
\slashed{k' } \gamma_5 - (k'^2 + c^2) \slashed{c} \gamma_5 }{(k'
-c)^2 (k'+c)^2} \right. \nonumber \\ && \left. \ \ \ \ \ \ \ \ \ \
\ \ \ \ \ \ \ \ \ \ \ \ \ \ \ \ \ \ \ \ \ \ \ \ \ \ \ \ \ \times
\gamma^\mu i \frac{ (k^2 + c^2 ) \slashed{k } - 2(k \cdot c)
\slashed{c} + 2(k \cdot c) \slashed{k } \gamma_5 - (k^2 + c^2)
\slashed{c} \gamma_5 }{(k -c)^2 (k+c)^2} \gamma^\rho   \right \}
\frac{-i \eta_{\nu \rho}}{(k-p)^2 } \nonumber \\ && \ \ \ \ \ \ \
\ \ \ \ \ \ \ = -i e^2 \int \frac{d^d k}{(2\pi)^d}
\frac{1}{(k-p)^2(k'-c)^2(k'+c)^2(k-c)^2(k+c)^2}\nonumber \\ && \
{} \ \ \ \ \ \ \ \ \ \ \ \ \ \ \ \ \ \ \ \ \ \ \ \ \ \ \ \ \ \ \ \
\ \ \ \ \ \ \times \gamma^\nu \{ ( k'^2 + c^2) \slashed{k'} - 2(k'
\cdot c) \slashed{c} +2 (k' \cdot c)\slashed{k'} \gamma_5 - (k'^2
+ c^2 ) \slashed{c} \gamma_5 \} \nonumber \\ && {} \ \ \ \ \ \ \ \
\ \ \ \ \ \ \ \ \ \ \ \ \ \ \ \ \ \ \ \ \ \ \ \ \ \ \ \ \ \ \
\times \gamma^\mu \{ ( k^2 + c^2) \slashed{k} - 2(k \cdot c)
\slashed{c} +2 (k \cdot c)\slashed{k} \gamma_5 - (k^2 + c^2 )
\slashed{c} \gamma_5 \} \gamma_\nu, \eqa where $ p^2 = p'^2 =
m_f^2 = 0 , \ k_\mu + q_\mu = k'_\mu, \ p_\mu + q_\mu = p'_\mu $.
Similarly, replacing the denominator with Feynman parameters and
taking only divergent terms, we obtain \bqa && \Gamma_1^\mu (p,
p', c) = -24ie^2 \int_{0}^{1} dx dy dz du dv \ \delta
(x+y+z+u+v-1) \int
\frac{d^d k}{(2 \pi)^d } \frac{1}{[l^2 - \Delta ]^5} \nonumber \\
&&  \times \gamma^\nu \{ (( l+a+q)^2 + c^2) ( \slashed{l} +
\slashed{a} + \slashed{q}) - 2( l \cdot c + a \cdot c + q \cdot
c)\slashed{c} + 2(l \cdot c + a \cdot c + q \cdot c )(\slashed{l}
+ \slashed{a} + \slashed{q} )\gamma_5 - ( (l+a+q)^2 + c^2 )
\slashed{c} \gamma_5 \} \nonumber \\ &&  \times \gamma^\mu \{ ((
l+a)^2 + c^2) ( \slashed{l} + \slashed{a} ) - 2( l \cdot c + a
\cdot c )\slashed{c} + 2(l \cdot c + a \cdot c )(\slashed{l} +
\slashed{a} )\gamma_5 - ( (l+a)^2 + c^2 ) \slashed{c} \gamma_5 \}
\gamma_\nu \nonumber \\ && = 48ie^2 \int_{0}^{1} dx dy dz du dv \
\delta (x+y+z+u+v-1) \left \{  \int \frac{d^d k}{(2 \pi)^d }
\frac{(l^2)^2 l_\lambda l_\rho}{[l^2 - \Delta ]^5} \right \} \{
 \gamma^\lambda \gamma^\mu \gamma^\rho
 \} + finite. \nonumber \\ && = \frac{e^2}{8 \pi^2 \epsilon} \gamma^\mu + finite, \eqa
where $ l_\mu = k_\mu - a_\mu, \ a_\mu = x p_\mu - (y+z) q_\mu + ( y-z+u-v)c_\mu .$

Third, we evaluate the vacuum polarization tensor. Although the
procedure is essentially the same as before, this calculation is
much more complicated, given by \bqa && i \Pi_1^{\mu \nu}(q,c)= -
(-ie)^2 \int \frac{d^d k}{(2 \pi)^d} Tr \left [ \gamma^\mu i
\frac{ (k^2+c^2)\slashed{k} -2(k \cdot c) \slashed{c} + 2(k \cdot
c)\slashed{k} \gamma_5 - (k^2 + c^2) \slashed{c} \gamma_5}{(k-c)^2
(k+c)^2} \right. \nonumber \\ && {} \ \left.  \ \ \ \ \ \ \ \ \ \
\ \ \ \ \ \ \ \ \ \ \ \ \ \ \ \ \ \ \times \gamma^\nu i
\frac{((k+q)^2 + c^2)(\slashed{k}+\slashed{q}) - 2(k \cdot c + q
\cdot c)\slashed{c}+2(k \cdot c + q \cdot
c)(\slashed{k}+\slashed{q})\gamma_5 -
((k+q)^2+c^2)\slashed{c}\gamma_5}{(k+q-c)^2 (k+q+c)^2 } \right ]
\nonumber \\ && \ \ \ \ \ \ \ \ \ \ \ \ \  = -e^2 \int \frac{d^d
k}{(2 \pi)^d } \frac{1}{(k-c)^2(k+c)^2(k+q-c)^2(k+q+c)^2}
\nonumber \\ && {} \ \ \ \ \ \ \ \ \ \ \ \ \ \ \ \ \ \ \ \ \ \ \ \
\ \ \times Tr \left [ \gamma^\mu \{ (k^2 + c^2)\slashed{k} - 2 ( k
\cdot c) \slashed{c} + 2(k \cdot c)\slashed{k} \gamma_5 - (k^2 +
c^2) \slashed{c} \gamma_5  \} \right. \nonumber \\ && \ \ \ \ \ \
\ \ \ \ \ \ \ \ \ \ \ \ \ \ \ \ \ \ \ \ \ \ \ \ \times \gamma^\nu
\{ ((k+q)^2 + c^2)(\slashed{k}+\slashed{q}) - 2(k \cdot c + q
\cdot c)\slashed{c}+2(k \cdot c + q \cdot
c)(\slashed{k}+\slashed{q})\gamma_5 -
((k+q)^2+c^2)\slashed{c}\gamma_5  \} \left.  \right ] \nonumber \\
&& \eqa Notice that the extra $(-1)$ factor comes from the fermion
loop in the diagram. Now, we have to change the loop momenta as
before. Straightforward but rather tedious algebras give us the
following expression \bqa && i \Pi_1^{\mu \nu} (q,c) = -6 e^2
\int_{0}^{1} dx dy dz \ \delta(x+y+z-1) \int \frac{d^d
l}{(2\pi)^d} \frac{1}{[l^2 - \Delta]^4} \nonumber \\ && \ \ \ \ \
\ \ \ \ \ \ \ \ \ \ \ \ \ \ \ \times [ \ \  (l^2)^3 \{ -4
\eta^{\mu \nu}  \} + 8(l^2)^2 l^\mu l^\nu \nonumber \\ && \ \ \ \
\ \ \ \ \ \ \ \ \ \ \ \ \ \ \ \ \ \ \ \ \ + (l^2)^2 \{ 4 a^\mu
b^\nu + 4 a^\nu b^\mu - 4 \eta^{\mu \nu} (a \cdot b) - 4i
\epsilon^{\mu \nu \rho \sigma} c_\rho q_\sigma + 8 c^\mu c^\nu -
12 \eta^{\mu \nu} c^2 - 4 \eta^{\mu \nu} (a^2 +b^2) \} \nonumber
\\ && \ \ \ \ \ \ \ \ \ \ \ \ \ \ \ \ \ \ \ \ \ \ \ \ \  + l^2
l^\mu l^\lambda \{ 8 h_\lambda h^\nu - 32 c_\lambda c^\nu \} + l^2
l^\nu l^\lambda \{ 8 h_\lambda h^\mu - 32 c_\lambda c^\mu \}
\nonumber \\ && \ \ \ \ \ \ \ \ \ \ \ \ \ \ \ \ \ \ \ \ \ \ \ \ \
+ l^2 l^\lambda l^\kappa \{ -8 h_\lambda h_\kappa \eta^{\mu \nu} -
16 a_\lambda b_\kappa \eta^{\mu \nu} + 16 c_\lambda c_\kappa
\eta^{\mu \nu} + 16i c_\lambda b_\sigma \eta_{\kappa \rho}
\epsilon^{\mu \nu \rho \sigma} + 16 i c_\lambda a_\rho
\eta_{\kappa \sigma} \epsilon^{\mu \nu \rho \sigma} \} \nonumber
\\ && \ \ \ \ \ \ \ \ \ \ \ \ \ \ \ \ \ \ \ \ \ \ \ \ \ + l^2
l^\mu l^\nu \{ 8a^2 + 8b^2 + 16 c^2 \} + l^\mu l^\nu l^\lambda
l^\kappa \{ 32 a_\lambda b_\kappa + 32 c_\kappa c_\lambda \} \ \ ]
\nonumber \\ && \ \ \ \ \ \ \ \ \ \ \ \ \ \ \ \ \ \ + finite,\eqa
where \bqa && l_\mu = k_\mu - a_\mu, \nonumber \\ && a_\mu = A
c_\mu + B q_\mu = (x-y+z-u) c_\mu + (u+z) q_\mu, \nonumber \\ &&
b_\mu = a_\mu + q_\mu = A c_\mu + (B+1) q_\mu, \nonumber \\ &&
h_\mu = a_\mu + b_\mu = 2A c_\mu + (2B+1)q_\mu, \nonumber \\ &&
\Delta = l^2 - [ x(k-c)^2 + y(k+c)^2 + z(k+q-c)^2 + u(k+q+c)^2 ]
\nonumber \\ && \ \ \ \  = 2(c \cdot q) \{ u^2 - u - xu + yu + z -
xz + zy - z^2 \} + q^2 \{ u^2 - u + z^2 - z + 2uz \} \nonumber \\
&& \ \ \ \ \ \ \ \ + c^2 \{ x^2 + y^2 + z^2 + u^2 - x - y - z - u
- 2xy + 2xz - 2yz - 2xu + 2yu - 2zu \}. \eqa Integrating over the
loop momenta $ l $, we get \bqa &&  i \Pi_1^{\mu \nu} (q,c) =
-\frac{6 e^2i}{\pi^2 \epsilon} \int_{0}^{1} dx dy dz \
\delta(x+y+z-1) \nonumber \\ && \times \left \{ - \eta^{\mu \nu}
\Delta +  c^\mu c^\nu ( \frac{10}{3}A^2 -\frac{2}{3}) + \eta^{\mu
\nu}c^2 (-\frac{7}{3}A^2 - \frac{1}{3}) + q^\mu q^\nu (
\frac{10}{3}B^2 + \frac{10}{3} B + \frac{1}{2}) + \eta^{\mu \nu}
q^2 (-\frac{7}{3}B^2 -\frac{7}{3}B-\frac{1}{2}) \right \} + finite
\nonumber \\ && = - \frac{e^2 i }{6 \pi^2 \epsilon} ( \eta^{\mu
\nu} q^2 - q^\mu q^\nu ) + finite. \eqa It is interesting to
observe that the divergent part is not modified by the background
chiral gauge field although it makes the expression much
complicated in the intermediate stage. Contributions from the
background chiral gauge field turn out to be cancelled exactly in
the polarization contribution. It is quite laborious to check this
cancellation.

We summarize leading divergent contributions of fermion one-loop
self-energy, one-loop gauge-fermion vertex, and gauge-boson
one-loop self-energy as follows \bqa && -i \Sigma_1(\slashed{p},
c) = \frac{e^2 i}{8 \pi^2 \epsilon}\slashed{p} + \frac{e^2 i}{8
\pi^2 \epsilon}\slashed{c} \gamma_5 + finite, \\ && \Gamma_1^\mu
(p,p', c) = \frac{e^2}{8\pi^2 \epsilon} \gamma^\mu + finite, \\ &&
i\Pi_1^{\mu \nu} (q, c) = - \frac{e^2 i}{6 \pi^2 \epsilon}
(\eta^{\mu \nu} q^2 - q^\mu q^\nu ) + finite. \eqa
%
%
%
As a result, we obtain
\bqa && \delta_\psi = - \frac{e^2}{8 \pi^2 \epsilon} + finite, \\
&& \delta_c = - \frac{e^2}{8 \pi^2 \epsilon} + finite, \\ &&
\delta_e = -\frac{e^2}{8\pi^2 \epsilon} + finite, \\ && \delta_A =
-\frac{e^2}{6\pi^2 \epsilon} + finite. \eqa

\subsection{Renormalization group equations}

We would like to point out that the Ward identity of $ Z_\psi =
Z_e $ is satisfied, guaranteeing the gauge invariance. We
emphasize that this serves quite a nontrivial check for our
renormalization group analysis, where complex dependencies for
chiral gauge fields are all cancelled to give rise to the Ward
identity. Recalling our perturbative analysis in the one-loop
level, the satisfaction of the Ward identity implies that the
renormalization group equation for the coupling constant does not
change, compared with the case in the absence of the background
chiral gauge field. On the other hand, the renormalization group
equation for the chiral gauge field is an essential point of our
study.

The beta function for the chiral gauge field is given by \bqa &&
\beta_{c_\nu} (\mu) = \mu \frac{d c_\nu}{d \mu} . \eqa Considering
that the bare quantity $ c_{B\nu} = Z_c c_\nu $ is independent of
the scale parameter of $ \mu $, we obtain the renormalization
group equation for the chiral gauge field \bqa && 0 = \frac{d}{d
\ln \mu} \ln c_{B\nu} = \frac{d M_c}{d e} \frac{ d e}{d \ln \mu} +
\frac{d}{d \ln \mu} \ln c_\nu , \eqa where \bqa && M_c = \ln Z_c =
\sum_{n=1}^{\infty} \frac{m_n (e, c)}{\epsilon^n} =
\frac{-\frac{e^2}{8 \pi^2 } + O(e^4)}{\epsilon} +
O(\frac{1}{\epsilon^2}) . \eqa In the one-loop level we obtain $
m_1 (e,c) = -\frac{e^2}{8\pi^2} $. Inserting $ \frac{d e}{d \ln
\mu} = \beta_e (\mu) - \epsilon e $ into the above expression, we
reach the following formula \bqa && 0 = \left ( (-\frac{e}{4
\pi^2}+ O(e^3)\ )\frac{1}{\epsilon} + O(\frac{1}{\epsilon^2} )
\right )( \beta_e (\mu ) - \epsilon e ) + \frac{1}{c_\nu} \beta_c
(\mu ) . \eqa Renormalizability guarantees the cancellation in
higher negative orders. As a result, we obtain \bqa &&
\beta_{c_\nu} ( \mu ) = \mu \frac{ d c_\nu}{d \mu} =
-\frac{e^2}{4\pi^2} c_\nu + O(e^4). \eqa

\subsection{Low temperature behaviors for background chiral gauge fields}

Solving the renormalization group equation for the coupling
constant \bqa && \beta_e (\mu) = \frac{d e}{d \ln \mu} =
\frac{e^3}{12 \pi^2 } , \eqa we obtain \bqa && e^2(\mu) =
\frac{e_D^2}{ 1 - \frac{e_D^2}{4 \pi^2} \ln \left ( \frac{\mu}{D}
\right ) } . \eqa

Substituting this solution into the renormalization group equation
for the chiral gauge field \bqa && \beta_{c_\nu} (\mu) = \frac{d
c_\nu}{d \ln \mu} = -\frac{e^2}{4\pi^2}c_\nu , \eqa we obtain \bqa
&& \ln \left ( \frac{c_\nu(\mu)}{c_\nu(D)} \right ) =-
\frac{1}{4\pi^2} \int_{D}^{\mu} d( \ln \mu ) \frac{e_D^2}{1-
\frac{e_D^2}{4 \pi^2 }\ln \left ( \frac{\mu}{D} \right ) }
\nonumber \\ && \ \ \ \ \ \ \ \ \ \ \ \ \ \ \ \ \ = - \int_{\ln
D}^{\ln \mu} dx \frac{1}{\left [ \alpha_D^{-1} + \ln D \right ] -
x} \nonumber \\ && \ \ \ \ \ \ \ \ \ \ \ \ \ \ \ \ \ =
\int_{0}^{\ln \frac{\mu}{D}} dx \frac{1}{ x - \alpha_D^{-1}}
\nonumber \\ && \ \ \ \ \ \ \ \ \ \ \ \ \ \ \ \ \ = \ln \left |
\frac{\ln \frac{\mu}{D} - \alpha_D^{-1} }{\alpha_D^{-1}} \right |
, \eqa where $ \alpha_D = \frac{e_D^2}{4 \pi^2} $ is the fine
structure constant at the cutoff scale. As a result, we find \bqa
&& c_\nu (\mu) = c_\nu (D) \ \alpha_D  \left | \ln \frac{\mu}{D} -
\alpha_D^{-1} \right |  . \eqa

\section{Quantum Boltzmann equation approach in the presence of both chiral anomaly and gauge interaction}

\subsection{A formal development of the quantum Boltzmann equation
in the presence of the topological $\boldsymbol{E}\cdot\boldsymbol{B}$ term}

Inserting the solutions [Eq. (11)] of semi-classical equations [Eq. (10)] 
into the quantum Boltzmann equation [Eq. (9)] and
performing some algebra, we obtain the following expression \bqa
&& \Bigl( 1 + \frac{e}{c} \boldsymbol{B} \cdot
\boldsymbol{\Omega}_{\boldsymbol{p}} \Bigr)^{-1} \frac{e}{c}
\boldsymbol{v}_{\boldsymbol{p}} \cdot \Bigl( \boldsymbol{B} \times
\frac{\partial G^{<}}{\partial \boldsymbol{p}} \Bigr) + \Bigl( 1 +
\frac{e}{c} \boldsymbol{B} \cdot
\boldsymbol{\Omega}_{\boldsymbol{p}} \Bigr)^{-2} \Bigl(
\frac{e}{c} \boldsymbol{v}_{\boldsymbol{p}} \times \boldsymbol{B}
\Bigr) \cdot ( e \boldsymbol{E} \times
\boldsymbol{\Omega}_{\boldsymbol{p}} ) \frac{\partial
G^{<}}{\partial \omega} \nn && - [A(\boldsymbol{p},\omega)]^{2}
\Bigl( - \frac{\partial f(\omega)}{\partial \omega} \Bigr) \Bigl(
1 + \frac{e}{c} \boldsymbol{B} \cdot
\boldsymbol{\Omega}_{\boldsymbol{p}} \Bigr)^{-2} \Bigl\{ e
\boldsymbol{E} + \frac{e^{2}}{c} (\boldsymbol{E} \cdot
\boldsymbol{B}) \boldsymbol{\Omega}_{\boldsymbol{p}} \Bigr\} \cdot
\Bigl\{ \boldsymbol{v}_{\boldsymbol{p}} + \frac{e}{c}
(\boldsymbol{\Omega}_{\boldsymbol{p}} \cdot
\boldsymbol{v}_{\boldsymbol{p}}) \boldsymbol{B} \Bigr\} \Gamma \nn
&& = - i [ 2 \Gamma G^{<} - \Sigma^{<} A ] , \eqa where the
argument of $(\boldsymbol{p},\omega)$ is omitted for simplicity.

The lesser self-energy is given by \cite{Kim_QBE} \bqa &&
\Sigma^{<}(\boldsymbol{p},\omega) = \sum_{\boldsymbol{q}}
\int_{0}^{\infty} d \nu \Bigl| \frac{\boldsymbol{p} \times
\boldsymbol{\hat{q}}}{m} \Bigr|^{2} \Im D_{a}(\boldsymbol{q},\nu)
\Bigl\{ [n(\nu) + 1]
G^{<}(\boldsymbol{p}+\boldsymbol{q},\omega+\nu) + n(\nu)
G^{<}(\boldsymbol{p}+\boldsymbol{q},\omega-\nu) \Bigr\} . \eqa
Here, we consider gauge interactions for example. Thus,
$D_{a}(\boldsymbol{q},\nu)$ represents the Green function of gauge
fluctuations. $n(\nu)$ is the Bose-Einstein distribution function.
One can replace the gauge-boson propagator with some other types
of fluctuations such as phonons, spin fluctuations, and etc. One
may consider the diffusion-mode propagator for weak
anti-localization, where the form of its vertex should be changed,
of course.

We write down the lesser Green's function in the following way
\cite{Kim_QBE} \bqa && G^{<}(\boldsymbol{p},\omega) = i
f(\omega) A(\boldsymbol{p},\omega) + i \Bigl( - \frac{\partial
f(\omega)}{\partial \omega} \Bigr) A(\boldsymbol{p},\omega)
\boldsymbol{v}_{\boldsymbol{p}} \cdot
\Lambda(\boldsymbol{p},\omega) , \eqa which consists of the
equilibrium part (the first term) and its correction term (the
second term). We call $\Lambda(\boldsymbol{p},\omega)$ ``vertex
distribution function" although it sounds somewhat confusing.
$f(\omega)$ is the Fermi-Dirac distribution function.

Inserting this ansatz into the quantum Boltzmann equation with the
expression of the lesser self-energy and performing some
straightforward algebra, we obtain \bqa && i \Bigl( 1 +
\frac{e}{c} \boldsymbol{B} \cdot
\boldsymbol{\Omega}_{\boldsymbol{p}} \Bigr)^{-1} \frac{e}{m c}
\boldsymbol{v}_{\boldsymbol{p}} \cdot \Bigl( \boldsymbol{B} \times
\frac{\partial \boldsymbol{p}_{\alpha}}{\partial \boldsymbol{p}}
\Bigr) \Lambda_{\alpha}(\boldsymbol{p},\omega) - i \Bigl( 1 +
\frac{e}{c} \boldsymbol{B} \cdot
\boldsymbol{\Omega}_{\boldsymbol{p}} \Bigr)^{-2} \Bigl(
\frac{e}{c} \boldsymbol{v}_{\boldsymbol{p}} \times \boldsymbol{B}
\Bigr) \cdot ( e \boldsymbol{E} \times
\boldsymbol{\Omega}_{\boldsymbol{p}} ) \nn && -
A(\boldsymbol{p},\omega) \Bigl( 1 + \frac{e}{c} \boldsymbol{B}
\cdot \boldsymbol{\Omega}_{\boldsymbol{p}} \Bigr)^{-2} \Bigl\{ e
\boldsymbol{E} + \frac{e^{2}}{c} (\boldsymbol{E} \cdot
\boldsymbol{B}) \boldsymbol{\Omega}_{\boldsymbol{p}} \Bigr\} \cdot
\Bigl\{ \boldsymbol{v}_{\boldsymbol{p}} + \frac{e}{c}
(\boldsymbol{\Omega}_{\boldsymbol{p}} \cdot
\boldsymbol{v}_{\boldsymbol{p}}) \boldsymbol{B} \Bigr\}
\Gamma(\boldsymbol{p},\omega) \nn && = 2
\Gamma(\boldsymbol{p},\omega) \boldsymbol{v}_{\boldsymbol{p}}
\cdot \Lambda(\boldsymbol{p},\omega) - \sum_{\boldsymbol{q}}
\int_{0}^{\infty} d \nu \Bigl| \frac{\boldsymbol{p} \times
\boldsymbol{\hat{q}}}{m} \Bigr|^{2} \Im D_{a}(\boldsymbol{q},\nu)
\Bigl\{ [n(\nu) + f(\omega+\nu)]
A(\boldsymbol{p}+\boldsymbol{q},\omega+\nu)
\boldsymbol{v}_{\boldsymbol{p}+\boldsymbol{q}} \cdot
\Lambda(\boldsymbol{p}+\boldsymbol{q},\omega+\nu) \nn && -
[n(-\nu) + f(\omega-\nu)]
A(\boldsymbol{p}+\boldsymbol{q},\omega-\nu)
\boldsymbol{v}_{\boldsymbol{p}+\boldsymbol{q}} \cdot
\Lambda(\boldsymbol{p}+\boldsymbol{q},\omega-\nu) \Bigr\} , \eqa
where we have used the following relation \bqa && 2
\Gamma(\boldsymbol{p},\omega) = \sum_{\boldsymbol{q}}
\int_{0}^{\infty} d \nu \Bigl| \frac{\boldsymbol{p} \times
\boldsymbol{\hat{q}}}{m} \Bigr|^{2} \Im D_{a}(\boldsymbol{q},\nu)
\Bigl\{ [ n(\nu) + f(\omega+\nu) ]
A(\boldsymbol{p}+\boldsymbol{q},\omega+\nu) - [ n(-\nu) + f(\omega
- \nu) ] A(\boldsymbol{p}+\boldsymbol{q},\omega-\nu) \Bigr\} . \nn
\eqa

Writing down the quantum Boltzmann equation in terms of components
and focusing on dynamics near the Fermi surface, we reach the
following expression for each component, \bqa &&
\frac{\Lambda_{F}^{x}(\omega)}{\tau_{tr}(\omega)} + i \Bigl( 1 +
\frac{e}{c} \boldsymbol{B} \cdot \boldsymbol{\Omega}_{F}
\Bigr)^{-1} \frac{e B_{z}}{m c} \Lambda_{F}^{y}(\omega) - i \Bigl(
1 + \frac{e}{c} \boldsymbol{B} \cdot \boldsymbol{\Omega}_{F}
\Bigr)^{-1} \frac{e B_{y}}{m c} \Lambda_{F}^{z}(\omega) \nn && = -
i \Bigl( 1 + \frac{e}{c} \boldsymbol{B} \cdot
\boldsymbol{\Omega}_{F} \Bigr)^{-2} \Bigl\{ \frac{e}{c} B_{y} ( e
\boldsymbol{E} \times \boldsymbol{\Omega}_{F} )_{z} - \frac{e}{c}
B_{z} ( e \boldsymbol{E} \times \boldsymbol{\Omega}_{F} )_{y}
\Bigr\} \nn && - A(\boldsymbol{p}_{F},\omega)
\Gamma(\boldsymbol{p}_{F},\omega) \Bigl( 1 + \frac{e}{c}
\boldsymbol{B} \cdot \boldsymbol{\Omega}_{F} \Bigr)^{-2} \Bigl\{ e
E_{x} + \frac{e^{2}}{c} (\boldsymbol{E} \cdot \boldsymbol{B})
\boldsymbol{\Omega}_{F}^{x} + \frac{e^{2}}{c} \Bigl( 1 +
\frac{e}{c} \boldsymbol{B} \cdot \boldsymbol{\Omega}_{F} \Bigr)
(\boldsymbol{E} \cdot \boldsymbol{B}) \boldsymbol{\Omega}_{F}^{x}
\Bigr\} , \eqa \bqa &&
\frac{\Lambda_{F}^{y}(\omega)}{\tau_{tr}(\omega)} - i \Bigl( 1 +
\frac{e}{c} \boldsymbol{B} \cdot \boldsymbol{\Omega}_{F}
\Bigr)^{-1} \frac{e B_{z} }{m c} \Lambda_{F}^{x}(\omega) + i
\Bigl( 1 + \frac{e}{c} \boldsymbol{B} \cdot
\boldsymbol{\Omega}_{F} \Bigr)^{-1} \frac{e B_{x}}{m c}
\Lambda_{F}^{z}(\omega) \nn && = - i \Bigl( 1 + \frac{e}{c}
\boldsymbol{B} \cdot \boldsymbol{\Omega}_{F} \Bigr)^{-2} \Bigl\{ -
\frac{e}{c} B_{x} ( e \boldsymbol{E} \times
\boldsymbol{\Omega}_{F} )_{z} + \frac{e}{c} B_{z} ( e
\boldsymbol{E} \times \boldsymbol{\Omega}_{F} )_{x} \Bigr\} \nn &&
- A(\boldsymbol{p}_{F},\omega) \Gamma(\boldsymbol{p}_{F},\omega)
\Bigl( 1 + \frac{e}{c} \boldsymbol{B} \cdot
\boldsymbol{\Omega}_{F} \Bigr)^{-2} \Bigl\{ e E_{y} +
\frac{e^{2}}{c} (\boldsymbol{E} \cdot \boldsymbol{B})
\boldsymbol{\Omega}_{F}^{y} + \frac{e^{2}}{c} \Bigl( 1 +
\frac{e}{c} \boldsymbol{B} \cdot \boldsymbol{\Omega}_{F} \Bigr)
(\boldsymbol{E} \cdot \boldsymbol{B}) \boldsymbol{\Omega}_{F}^{y}
\Bigr\} , \eqa and \bqa &&
\frac{\Lambda_{F}^{z}(\omega)}{\tau_{tr}(\omega)} + i \Bigl( 1 +
\frac{e}{c} \boldsymbol{B} \cdot \boldsymbol{\Omega}_{F}
\Bigr)^{-1} \frac{e B_{y} }{m c} \Lambda_{F}^{x}(\omega) - i
\Bigl( 1 + \frac{e}{c} \boldsymbol{B} \cdot
\boldsymbol{\Omega}_{F} \Bigr)^{-1} \frac{e B_{x}}{m c}
\Lambda_{F}^{y}(\omega) \nn && = - i \Bigl( 1 + \frac{e}{c}
\boldsymbol{B} \cdot \boldsymbol{\Omega}_{F} \Bigr)^{-2} \Bigl\{ -
\frac{e}{c} B_{y} ( e \boldsymbol{E} \times
\boldsymbol{\Omega}_{F} )_{x} + \frac{e}{c} B_{x} ( e
\boldsymbol{E} \times \boldsymbol{\Omega}_{F} )_{y} \Bigr\} \nn &&
- A(\boldsymbol{p}_{F},\omega) \Gamma(\boldsymbol{p}_{F},\omega)
\Bigl( 1 + \frac{e}{c} \boldsymbol{B} \cdot
\boldsymbol{\Omega}_{F} \Bigr)^{-2} \Bigl\{ e E_{z} +
\frac{e^{2}}{c} (\boldsymbol{E} \cdot \boldsymbol{B})
\boldsymbol{\Omega}_{F}^{z} + \frac{e^{2}}{c} \Bigl( 1 +
\frac{e}{c} \boldsymbol{B} \cdot \boldsymbol{\Omega}_{F} \Bigr)
(\boldsymbol{E} \cdot \boldsymbol{B}) \boldsymbol{\Omega}_{F}^{z}
\Bigr\} , \eqa where the transport time is given by \bqa &&
\frac{1}{\tau_{tr}(\omega)} = \sum_{\boldsymbol{q}}
\int_{0}^{\infty} d \nu \Bigl| \frac{\boldsymbol{p}_{F} \times
\boldsymbol{\hat{q}}}{m} \Bigr|^{2} \Im D_{a}(\boldsymbol{q},\nu)
(1 -  \cos \theta) \Bigl\{ [n(\nu) + f(\omega+\nu)]
A(\boldsymbol{p}_{F}+\boldsymbol{q},\omega+\nu) \nn && - [n(-\nu)
+ f(\omega-\nu)] A(\boldsymbol{p}_{F}+\boldsymbol{q},\omega-\nu)
\Bigr\} . \eqa We note the $1 - \cos \theta$ factor in this
expression, which extracts out back scattering contributions.

\subsection{Current formulation}

It is natural to define a current in the following way
\cite{DTSon_Boltzmann} \bqa && \boldsymbol{J} = - e
\frac{1}{\beta} \sum_{i\omega} \int \frac{d^{3}
\boldsymbol{p}}{(2\pi \hbar)^{3}} \Bigl( 1 + \frac{e}{c}
\boldsymbol{B} \cdot \boldsymbol{\Omega}_{\boldsymbol{p}}
\Bigr)^{-1} \Bigl\{ \boldsymbol{v}_{\boldsymbol{p}} + e
\boldsymbol{E} \times \boldsymbol{\Omega}_{\boldsymbol{p}} +
\frac{e}{c} (\boldsymbol{\Omega}_{\boldsymbol{p}} \cdot
\boldsymbol{v}_{\boldsymbol{p}}) \boldsymbol{B} \Bigr\}[ - i
G^{<}(\boldsymbol{p},i\omega)] . \eqa We note the
$\dot{\boldsymbol{r}}$ term in the integral expression.

Inserting the ansatz for the lesser Green's function into the
above expression, we obtain \bqa && \boldsymbol{J} = - e^{2}
\frac{1}{\beta} \sum_{i\omega} \int \frac{d^{3}
\boldsymbol{p}}{(2\pi \hbar)^{3}} \Bigl( 1 + \frac{e}{c}
\boldsymbol{B} \cdot \boldsymbol{\Omega}_{\boldsymbol{p}}
\Bigr)^{-1} (\boldsymbol{E} \times
\boldsymbol{\Omega}_{\boldsymbol{p}}) f(\omega)
A(\boldsymbol{p},\omega) \nn && - e \frac{1}{\beta} \sum_{i\omega}
\int \frac{d^{3} \boldsymbol{p}}{(2\pi \hbar)^{3}} \Bigl( 1 +
\frac{e}{c} \boldsymbol{B} \cdot
\boldsymbol{\Omega}_{\boldsymbol{p}} \Bigr)^{-1} \Bigl\{
\boldsymbol{v}_{\boldsymbol{p}} + \frac{e}{c}
(\boldsymbol{\Omega}_{\boldsymbol{p}} \cdot
\boldsymbol{v}_{\boldsymbol{p}}) \boldsymbol{B} \Bigr\} \Bigl( -
\frac{\partial f(\omega)}{\partial \omega} \Bigr)
A(\boldsymbol{p},\omega) \boldsymbol{v}_{\boldsymbol{p}} \cdot
\Lambda(\boldsymbol{p},\omega) . \eqa Then, each component is
given by \bqa && J_{x} = - e^{2} \frac{1}{\beta} \sum_{i\omega}
\int \frac{d^{3} \boldsymbol{p}}{(2\pi \hbar)^{3}} \Bigl( 1 +
\frac{e}{c} \boldsymbol{B} \cdot
\boldsymbol{\Omega}_{\boldsymbol{p}} \Bigr)^{-1} (E_{y}
\boldsymbol{\Omega}_{\boldsymbol{p}}^{z} - E_{z}
\boldsymbol{\Omega}_{\boldsymbol{p}}^{y}) f(\omega)
A(\boldsymbol{p},\omega) \nn && - e \frac{1}{\beta} \sum_{i\omega}
\int \frac{d^{3} \boldsymbol{p}}{(2\pi \hbar)^{3}} \Bigl( 1 +
\frac{e}{c} \boldsymbol{B} \cdot
\boldsymbol{\Omega}_{\boldsymbol{p}} \Bigr)^{-1} (v_{F}^{x})^{2}
\Bigl( - \frac{\partial f(\omega)}{\partial \omega} \Bigr)
A(\boldsymbol{p},\omega) \Lambda_{x}(\boldsymbol{p},\omega) \nn &&
- e \frac{1}{\beta} \sum_{i\omega} \int \frac{d^{3}
\boldsymbol{p}}{(2\pi \hbar)^{3}} \Bigl( 1 + \frac{e}{c}
\boldsymbol{B} \cdot \boldsymbol{\Omega}_{\boldsymbol{p}}
\Bigr)^{-1} \frac{e}{c} B_{x} \Bigl\{ (v_{\boldsymbol{p}}^{x})^{2}
\boldsymbol{\Omega}_{\boldsymbol{p}}^{x}
\Lambda_{x}(\boldsymbol{p},\omega) + (v_{\boldsymbol{p}}^{y})^{2}
\boldsymbol{\Omega}_{\boldsymbol{p}}^{y}
\Lambda_{y}(\boldsymbol{p},\omega) + (v_{\boldsymbol{p}}^{z})^{2}
\boldsymbol{\Omega}_{\boldsymbol{p}}^{z}
\Lambda_{z}(\boldsymbol{p},\omega) \Bigr\} \Bigl( - \frac{\partial
f(\omega)}{\partial \omega} \Bigr) A(\boldsymbol{p},\omega) , \nn
\eqa and \bqa && J_{y} = e^{2} \frac{1}{\beta} \sum_{i\omega} \int
\frac{d^{3} \boldsymbol{p}}{(2\pi \hbar)^{3}} \Bigl( 1 +
\frac{e}{c} \boldsymbol{B} \cdot
\boldsymbol{\Omega}_{\boldsymbol{p}} \Bigr)^{-1} (E_{z}
\boldsymbol{\Omega}_{\boldsymbol{p}}^{x} - E_{x}
\boldsymbol{\Omega}_{\boldsymbol{p}}^{z}) f(\omega)
A(\boldsymbol{p},\omega) \nn && + e \frac{1}{\beta} \sum_{i\omega}
\int \frac{d^{3} \boldsymbol{p}}{(2\pi \hbar)^{3}} \Bigl( 1 +
\frac{e}{c} \boldsymbol{B} \cdot
\boldsymbol{\Omega}_{\boldsymbol{p}} \Bigr)^{-1} (v_{F}^{y})^{2}
\Bigl( - \frac{\partial f(\omega)}{\partial \omega} \Bigr)
A(\boldsymbol{p},\omega) \Lambda_{y}(\boldsymbol{p},\omega) \nn &&
+ e \frac{1}{\beta} \sum_{i\omega} \int \frac{d^{3}
\boldsymbol{p}}{(2\pi \hbar)^{3}} \Bigl( 1 + \frac{e}{c}
\boldsymbol{B} \cdot \boldsymbol{\Omega}_{\boldsymbol{p}}
\Bigr)^{-1} \frac{e}{c} B_{y} \Bigl\{ (v_{\boldsymbol{p}}^{x})^{2}
\boldsymbol{\Omega}_{\boldsymbol{p}}^{x}
\Lambda_{x}(\boldsymbol{p},\omega) + (v_{\boldsymbol{p}}^{y})^{2}
\boldsymbol{\Omega}_{\boldsymbol{p}}^{y}
\Lambda_{y}(\boldsymbol{p},\omega) + (v_{\boldsymbol{p}}^{z})^{2}
\boldsymbol{\Omega}_{\boldsymbol{p}}^{z}
\Lambda_{z}(\boldsymbol{p},\omega) \Bigr\} \Bigl( - \frac{\partial
f(\omega)}{\partial \omega} \Bigr) A(\boldsymbol{p},\omega) . \nn
\eqa

\subsection{Longitudinal magneto-transport}

Solving the quantum Boltzmann equation in the unconventional setup
of $\boldsymbol{B} = B_{x} \boldsymbol{\hat{x}}$ and
$\boldsymbol{E} = E_{x} \boldsymbol{\hat{x}}$, we find \bqa &&
\Lambda_{F}^{x}(\omega) = - e A(\boldsymbol{p}_{F},\omega)
\frac{\tau_{tr}(\omega)}{\tau_{sc}(\omega)} E_{x} , \eqa \bqa &&
\Lambda_{F}^{y}(\omega) = m e \frac{\omega_{c}^{x}
\tau_{tr}(\omega)}{\Bigl(1 + \frac{e}{c} B_{x}
\boldsymbol{\Omega}_{F}^{x}\Bigr)^{2} + [\omega_{c}^{x}
\tau_{tr}(\omega)]^{2}} \Bigl( - \boldsymbol{\Omega}_{F}^{z} +
\boldsymbol{\Omega}_{F}^{y} \frac{\omega_{c}^{x}
\tau_{tr}(\omega)}{1 + \frac{e}{c} B_{x}
\boldsymbol{\Omega}_{F}^{x}} \Bigr) E_{x} \nn && -
A(\boldsymbol{p}_{F},\omega) \frac{
\frac{\tau_{tr}(\omega)}{\tau_{sc}(\omega)}}{\Bigl(1 + \frac{e}{c}
B_{x} \boldsymbol{\Omega}_{F}^{x}\Bigr)^{2} + [\omega_{c}^{x}
\tau_{tr}(\omega)]^{2}} \Bigl\{ \frac{e^{2}}{c} + \frac{e^{2}}{c}
\Bigl( 1 + \frac{e}{c} B_{x} \boldsymbol{\Omega}_{F}^{x} \Bigr)
\Bigr\} \Bigl( \boldsymbol{\Omega}_{F}^{y} +
\boldsymbol{\Omega}_{F}^{z} \frac{\omega_{c}^{x}
\tau_{tr}(\omega)}{1 + \frac{e}{c} B_{x}
\boldsymbol{\Omega}_{F}^{x}} \Bigr) E_{x} B_{x} , \eqa and \bqa &&
\Lambda_{F}^{z}(\omega) = m e \frac{\omega_{c}^{x}
\tau_{tr}(\omega)}{\Bigl(1 + \frac{e}{c} B_{x}
\boldsymbol{\Omega}_{F}^{x}\Bigr)^{2} + [\omega_{c}^{x}
\tau_{tr}(\omega)]^{2}} \Bigl( \boldsymbol{\Omega}_{F}^{y} +
\boldsymbol{\Omega}_{F}^{z} \frac{\omega_{c}^{x}
\tau_{tr}(\omega)}{1 + \frac{e}{c} B_{x}
\boldsymbol{\Omega}_{F}^{x}} \Bigr) E_{x} \nn && -
A(\boldsymbol{p}_{F},\omega) \frac{
\frac{\tau_{tr}(\omega)}{\tau_{sc}(\omega)}}{\Bigl(1 + \frac{e}{c}
B_{x} \boldsymbol{\Omega}_{F}^{x}\Bigr)^{2} + [\omega_{c}^{x}
\tau_{tr}(\omega)]^{2}} \Bigl\{ \frac{e^{2}}{c} + \frac{e^{2}}{c}
\Bigl( 1 + \frac{e}{c} B_{x} \boldsymbol{\Omega}_{F}^{x} \Bigr)
\Bigr\} \Bigl( \boldsymbol{\Omega}_{F}^{z} -
\boldsymbol{\Omega}_{F}^{y} \frac{\omega_{c}^{x}
\tau_{tr}(\omega)}{1 + \frac{e}{c} B_{x}
\boldsymbol{\Omega}_{F}^{x}} \Bigr) E_{x} B_{x} , \eqa where
$\omega_{c}^{x} = \frac{e B_{x}}{m c}$ is the ``cyclotron"
frequency associated with the $B_{x}$ field. We notice that there
are $\boldsymbol{E}\cdot\boldsymbol{B} = E_{x} B_{x}$ terms, which
are topological in their origin.

Inserting these vertex distribution functions into the current
formula, we obtain a rather complicated expression for the $x-$
component of the current, \bqa && J_{x} = e^{2} \frac{1}{\beta}
\sum_{i\omega} \int \frac{d^{3} \boldsymbol{p}}{(2\pi \hbar)^{3}}
\frac{1}{1 + \frac{e}{c} B_{x}\boldsymbol{\Omega}_{F}^{x}}
(v_{F}^{x})^{2} \Bigl( - \frac{\partial f(\omega)}{\partial
\omega} \Bigr) [A(\boldsymbol{p}_{F},\omega)]^{2}
\frac{\tau_{tr}(\omega)}{\tau_{sc}(\omega)} E_{x} \nn && +
\frac{e^{3}}{c} \frac{1}{\beta} \sum_{i\omega} \int \frac{d^{3}
\boldsymbol{p}}{(2\pi \hbar)^{3}} \frac{1}{1 + \frac{e}{c}
B_{x}\boldsymbol{\Omega}_{F}^{x}} (v_{\boldsymbol{p}}^{x})^{2}
\boldsymbol{\Omega}_{F}^{x} \Bigl( - \frac{\partial
f(\omega)}{\partial \omega} \Bigr)
[A(\boldsymbol{p}_{F},\omega)]^{2}
\frac{\tau_{tr}(\omega)}{\tau_{sc}(\omega)} B_{x} E_{x} \nn && -
\frac{m e^{3}}{c} \frac{1}{\beta} \sum_{i\omega} \int \frac{d^{3}
\boldsymbol{p}}{(2\pi \hbar)^{3}} \frac{1}{1 + \frac{e}{c} B_{x}
\boldsymbol{\Omega}_{F}^{x}} (v_{F}^{y})^{2} \Bigl( -
\frac{\partial f(\omega)}{\partial \omega} \Bigr)
A(\boldsymbol{p}_{F},\omega) \nn && \frac{\omega_{c}^{x}
\tau_{tr}(\omega)}{\Bigl(1 + \frac{e}{c} B_{x}
\boldsymbol{\Omega}_{F}^{x}\Bigr)^{2} + [\omega_{c}^{x}
\tau_{tr}(\omega)]^{2}} \Bigl( - \boldsymbol{\Omega}_{F}^{y}
\boldsymbol{\Omega}_{F}^{z} + (\boldsymbol{\Omega}_{F}^{y})^{2}
\frac{\omega_{c}^{x} \tau_{tr}(\omega)}{1 + \frac{e}{c} B_{x}
\boldsymbol{\Omega}_{F}^{x}} \Bigr) B_{x} E_{x} \nn && +
\frac{e^{4}}{c^{2}} \frac{1}{\beta} \sum_{i\omega} \int
\frac{d^{3} \boldsymbol{p}}{(2\pi \hbar)^{3}} \frac{2 +
\frac{e}{c} B_{x} \boldsymbol{\Omega}_{F}^{x}}{1 + \frac{e}{c}
B_{x}\boldsymbol{\Omega}_{F}^{x}} (v_{F}^{y})^{2} \Bigl( -
\frac{\partial f(\omega)}{\partial \omega} \Bigr)
[A(\boldsymbol{p}_{F},\omega)]^{2} \frac{
\frac{\tau_{tr}(\omega)}{\tau_{sc}(\omega)} \Bigl(
(\boldsymbol{\Omega}_{F}^{y})^{2} + \boldsymbol{\Omega}_{F}^{y}
\boldsymbol{\Omega}_{F}^{z} \frac{\omega_{c}^{x}
\tau_{tr}(\omega)}{1 + \frac{e}{c} B_{x}
\boldsymbol{\Omega}_{F}^{x}} \Bigr)}{\Bigl(1 + \frac{e}{c} B_{x}
\boldsymbol{\Omega}_{F}^{x}\Bigr)^{2} + [\omega_{c}^{x}
\tau_{tr}(\omega)]^{2}} E_{x} B_{x}^{2} \nn && - \frac{m e^{3}}{c}
\frac{1}{\beta} \sum_{i\omega} \int \frac{d^{3}
\boldsymbol{p}}{(2\pi \hbar)^{3}} \frac{1}{1 + \frac{e}{c} B_{x}
\boldsymbol{\Omega}_{F}^{x}} (v_{F}^{z})^{2} \Bigl( -
\frac{\partial f(\omega)}{\partial \omega} \Bigr)
A(\boldsymbol{p}_{F},\omega) \nn && \frac{\omega_{c}^{x}
\tau_{tr}(\omega)}{\Bigl(1 + \frac{e}{c} B_{x}
\boldsymbol{\Omega}_{F}^{x}\Bigr)^{2} + [\omega_{c}^{x}
\tau_{tr}(\omega)]^{2}} \Bigl( \boldsymbol{\Omega}_{F}^{y}
\boldsymbol{\Omega}_{F}^{z} + (\boldsymbol{\Omega}_{F}^{z})^{2}
\frac{\omega_{c}^{x} \tau_{tr}(\omega)}{1 + \frac{e}{c} B_{x}
\boldsymbol{\Omega}_{F}^{x}} \Bigr) B_{x} E_{x} \nn && +
\frac{e^{4}}{c^{2}} \frac{1}{\beta} \sum_{i\omega} \int
\frac{d^{3} \boldsymbol{p}}{(2\pi \hbar)^{3}} \frac{2 +
\frac{e}{c} B_{x} \boldsymbol{\Omega}_{F}^{x}}{1 + \frac{e}{c}
B_{x} \boldsymbol{\Omega}_{F}^{x}} (v_{F}^{z})^{2} \Bigl( -
\frac{\partial f(\omega)}{\partial \omega} \Bigr)
[A(\boldsymbol{p}_{F},\omega)]^{2} \frac{
\frac{\tau_{tr}(\omega)}{\tau_{sc}(\omega)} \Bigl(
(\boldsymbol{\Omega}_{F}^{z})^{2} - \boldsymbol{\Omega}_{F}^{y}
\boldsymbol{\Omega}_{F}^{z} \frac{\omega_{c}^{x}
\tau_{tr}(\omega)}{1 + \frac{e}{c} B_{x}
\boldsymbol{\Omega}_{F}^{x}} \Bigr)}{\Bigl(1 + \frac{e}{c} B_{x}
\boldsymbol{\Omega}_{F}^{x}\Bigr)^{2} + [\omega_{c}^{x}
\tau_{tr}(\omega)]^{2}} E_{x} B_{x}^{2} . \eqa

Expanding the above expression up to the second order for the
Berry curvature and keeping only even-power contributions
\cite{DTSon_Boltzmann}, we obtain \bqa && J_{x} \approx e^{2}
\frac{1}{\beta} \sum_{i\omega} \int \frac{d^{3}
\boldsymbol{p}}{(2\pi \hbar)^{3}} (v_{F}^{x})^{2} \Bigl( -
\frac{\partial f(\omega)}{\partial \omega} \Bigr)
[A(\boldsymbol{p}_{F},\omega)]^{2}
\frac{\tau_{tr}(\omega)}{\tau_{sc}(\omega)} E_{x} \nn && - 2
\frac{m e^{3}}{c} \frac{1}{\beta} \sum_{i\omega} \int \frac{d^{3}
\boldsymbol{p}}{(2\pi \hbar)^{3}} (v_{F}^{y})^{2} \Bigl( -
\frac{\partial f(\omega)}{\partial \omega} \Bigr)
A(\boldsymbol{p}_{F},\omega) \frac{
(\boldsymbol{\Omega}_{F}^{y})^{2} \omega_{c}^{x}
\tau_{tr}(\omega)}{1 + [\omega_{c}^{x} \tau_{tr}(\omega)]^{2}}
[\omega_{c}^{x} \tau_{tr}(\omega)] B_{x} E_{x} \nn && + 2
\frac{e^{4}}{c^{2}} \frac{1}{\beta} \sum_{i\omega} \int
\frac{d^{3} \boldsymbol{p}}{(2\pi \hbar)^{3}} (v_{F}^{y})^{2}
\Bigl( - \frac{\partial f(\omega)}{\partial \omega} \Bigr)
[A(\boldsymbol{p}_{F},\omega)]^{2} \frac{
\frac{\tau_{tr}(\omega)}{\tau_{sc}(\omega)}
(\boldsymbol{\Omega}_{F}^{y})^{2} }{1 + [\omega_{c}^{x}
\tau_{tr}(\omega)]^{2}} E_{x} B_{x}^{2} \nn && = \mathcal{C} N_{F}
e^{2} v_{F}^{2} \tau_{tr}(T) E_{x} + 2 \mathcal{C}'
\frac{e^{4}}{c^{2}} N_{F} v_{F}^{2} \frac{ \tau_{tr}(T)}{1 +
[\omega_{c}^{x} \tau_{tr}(T)]^{2}} B_{x}^{2} E_{x} - 2
\mathcal{C}'' \frac{m e^{3}}{c} N_{F} v_{F}^{2} \frac{\tau_{sc}(T)
[\omega_{c}^{x} \tau_{tr}(T)]^{2}}{1 + [\omega_{c}^{x}
\tau_{tr}(T)]^{2}} B_{x} E_{x} . \eqa The first term is also the
conventional contribution near the Fermi surface, but there is no
dependence for magnetic fields. This is certainly expected because
the magnetic field is in the same direction as the electric field.
On the other hand, the second contribution originates from the
topological $\boldsymbol{E}\cdot\boldsymbol{B}$ term. The third
term is also anomalous, which results from the Berry curvature but
not from the $\boldsymbol{E}\cdot\boldsymbol{B}$ term.

\subsection{Discussion}

The longitudinal magnetoconductivity is \bqa &&
\sigma_{L}(B_{x},T) = \mathcal{C} N_{F} e^{2} v_{F}^{2}
\tau_{tr}(T) + 2 \mathcal{C}' \frac{e^{4}}{c^{2}} N_{F} v_{F}^{2}
\frac{ \tau_{tr}(T)}{1 + [\omega_{c}^{x} \tau_{tr}(T)]^{2}}
B_{x}^{2} - 2 \mathcal{C}'' \frac{m e^{3}}{c} N_{F} v_{F}^{2}
\frac{\tau_{sc}(T) [\omega_{c}^{x} \tau_{tr}(T)]^{2}}{1 +
[\omega_{c}^{x} \tau_{tr}(T)]^{2}} B_{x} . \eqa If we limit our
discussion on low magnetic fields, we are allowed to neglect the
last contribution. Then, the above expression can be rewritten as
follows \bqa && \sigma_{L}(B_{x},T) = (1 + \mathcal{C}_{W}
B_{x}^{2}) \sigma_{n}(T) , \eqa where $\sigma_{n}(T) = \mathcal{C}
N_{F} e^{2} v_{F}^{2} \tau_{tr}(T)$ is the normal conductivity and
$\mathcal{C}_{W} = 2 (\mathcal{C}' / \mathcal{C}) (e^{2}/c^{2})$
is a positive constant.

Our proposal is to replace the $B_{x}^{2}$ term with $[c(T)]^{2}$,
where $c(T)$ represents the distance between two Weyl points.
Then, the final expression for the ``longitudinal" conductivity
becomes \bqa && \sigma_{L}(T) \longrightarrow (1 + \mathcal{K}
[c(T)]^{2}) \sigma_{n}(T) , \eqa where $\mathcal{K}$ is a positive
numerical constant and the normal conductivity is determined by
intra-scattering events at one Weyl point.

\end{widetext}


\begin{thebibliography}{9}
\bibitem{Peskin} M. E. Peskin and D. V. Schroeder, {\it An Introduction to Quantum Field
Theory}, Ch. 19 (Addison-Wesley Publishing Company, Seoul, 1997)
\bibitem{GSW_String} M. B. Green, J. H. Schwarz, and E. Witten,
{\it Superstring theory} vol. 1 {\it Introduction} Ch. 3
(Cambridge Univeristy Press, New York, 1987); M. B. Green, J. H.
Schwarz, and E. Witten, {\it Superstring theory} vol. 2 {\it Loop
amplitudes, Anomalies, and Phenomenology} Ch. 10 and Ch. 13
(Cambridge Univeristy Press, New York, 2007).
\bibitem{Fendley_NLsM} P. Fendley, arXiv:cond-mat/0006360,
Lecture at the NATO Advanced Study Institute/EC Summer School
on New Theoretical Approaches to Strongly Correlated Systems,
Newton Institute, Cambridge, UK, April 10-20, 2000.
\bibitem{Review_FZM} A. J. Niemi and G.W. Semenoff, Physics Reports {\bf 135}, 99
(1986).
\bibitem{Senthil} T. Senthil, A. Vishwanath, L. Balents, S.
Sachdev, and M. P. A. Fisher, Science {\bf 303}, 1490 (2004); T.
Senthil, L. Balents, S. Sachdev, A. Vishwanath, and M. P.A.
Fisher, Phys. Rev. B {\bf 70}, 144407 (2004).
\bibitem{Tanaka} A. Tanaka and X. Hu, Phys. Rev. Lett. {\bf
95}, 036402 (2005); A. Tanaka and X. Hu, Phys. Rev. Lett. {\bf
88}, 127004 (2002).
\bibitem{Review_AHE} D. Xiao, M.-C. Chang, and Q. Niu, Rev. Mod. Phys. {\bf 82},
1959 (2010); N. Nagaosa, J. Sinova, S. Onoda, A. H. MacDonald, and
N. P. Ong, Rev. Mod. Phys. {\bf 82}, 1539 (2010).
\bibitem{Review_TI} M. Z. Hasan and C. L. Kane, Rev. Mod. Phys. {\bf 82}, 3045
(2010); X.-L. Qi and S.-C. Zhang, Rev. Mod. Phys. {\bf 83}, 1057
(2011).
\bibitem{Kim_AdS5} K.-S. Kim and T. Tsukioka, Phys. Rev. D {\bf 85}, 045011
(2012).
\bibitem{Weyl_Metal} X. Wan, A. M. Turner, A. Vishwanath, and S. Y. Savrasov, Phys. Rev. B {\bf 83}, 205101 (2011);
A. A. Burkov and L. Balents, Phys. Rev. Lett. {\bf 107}, 127205
(2011); G. Xu, H. Weng, Z. Wang, X. Dai, and Z. Fang, Phys. Rev.
Lett. {\bf 107}, 186806 (2011); A. A. Burkov, M. D. Hook, and L.
Balents, Phys. Rev. B {\bf 84}, 235126 (2011); P. Hosur, S. A.
Parameswaran, and A. Vishwanath, Phys. Rev. Lett. {\bf 108},
046602 (2012); K.-Y. Yang, Y.-M. Lu, and Y. Ran, Phys. Rev. B {\bf
84}, 075129 (2011); C. Fang, M. J. Gilbert, X. Dai, and B. A.
Bernevig, Phys. Rev. Lett. {\bf 108}, 266802 (2012).
\bibitem{Anomaly_Weyl} H. B. Nielsen and M. Ninomiya, Phys. Lett. {\bf 130B}, 390 (1983).
\bibitem{DTSon_Boltzmann} D. T. Son and B. Z. Spivak,
arXiv:1206.1627.
\bibitem{Graphene_Review} A. H. Castro Neto, F. Guinea, N. M. R. Peres, K. S. Novoselov, and A. K.
Geim, Rev. Mod. Phys. {\bf 81}, 109 (2009); S. D. Sarma, S. Adam,
E. H. Hwang, and E. Rossi, Rev. Mod. Phys. {\bf 83}, 407 (2011);
D. S. L. Abergel, V. Apalkov, J. Berashevich, K. Ziegler, and T.
Chakraborty, Advances in Physics, {\bf 59 (4)}, 261 (2010); E. R.
Mucciolo and C. H. Lewenkopf, J. Phys.: Condens. Matter {\bf 22},
273201 (2010).
\bibitem{Nagaosa_Lee_Wen} P. A. Lee, N. Nagaosa, and X.-G. Wen, Rev. Mod. Phys. {\bf 78}, 17
(2006).
\bibitem{Fujikawa_Method} P. Goswami and S. Tewari,
arXiv:1210.6352.
\bibitem{Ran_AHE_Weyl} K.-Y. Yang, Y.-M. Lu, and Y. Ran, Phys. Rev. B {\bf 84}, 075129 (2011).
\bibitem{Chiral_Gauge_Field} C.-X. Liu, P. Ye, and X.-L. Qi,
arXiv:1204.6551.
\bibitem{TSL_YB} S. Bhattacharjee, Y.-B. Kim, S.-S. Lee, and D.-H. Lee, Phys. Rev. B {\bf 85},
224428 (2012).
\bibitem{Axion_Insulator} J. Wang, R. Li, S.-C. Zhang, and X.-L. Qi, Phys. Rev. Lett. {\bf 106}, 126403
(2011). 
\bibitem{MI_TI_Kim} H.-J. Kim, K.-S. Kim, J.-F. Wang, V. A. Kulbachinskii, K. Ogawa,
M. Sasaki, A. Ohnishi, M. Kitaura, Y.-Y. Wu, L. Li, I. Yamamoto,
J. Azuma, and M. Kamada, Phys. Rev. Lett. {\bf 110}, 136601 (2013).
\bibitem{Kim_QBE} K.-S. Kim, Phys. Rev. B {\bf 84}, 085117 (2011); K.-S. Kim, Phys. Rev. B {\bf 84}, 085111 (2011); 
K.-S. Kim and C. Pepin, J. Phys.: Condens. Matter {\bf 22}, 025601 (2010);
K.-S. Kim and C. Pepin, Phys. Rev. Lett. {\bf 102}, 156404 (2009)
\bibitem{Matrix_Boltzmann_Equation} A. A. Kovalev, Y. Tserkovnyak, K. Vyborny, and J.
Sinova, Phys. Rev. B {\bf 79}, 195129 (2009).
\bibitem{Shindou_QBE} R. Shindou and L. Balents, Phys. Rev. B {\bf 77}, 035110 (2008).
\bibitem{Mahan_Boltzmann} G. D. Mahan, {\it Many-Particle Physics}, 3rd ed. (Kluwer Academic/Plenum, New York, 2000). 
\bibitem{Weyl_LMR} H.-J. Kim, K.-S. Kim, J.-F. Wang, M. Sasaki,
N. Satoh, A. Ohnishi, M. Kitaura, M. Yang, and L. Li, ``Dirac vs.
Weyl in topological insulators: Adler-Bell-Jackiw anomaly in
transport phenomena," submitted to Nature Phys. 
\end{thebibliography}
\end{document}